\documentclass[english,prd,aps, nofootinbib,superscriptaddress,amsfonts]{revtex4-1}
\usepackage[T1]{fontenc}
\usepackage[latin9]{inputenc}
\usepackage{babel}
\usepackage{verbatim}
\usepackage{amsmath}
\usepackage{amssymb}
\usepackage{esint}
\usepackage[unicode=true,pdfusetitle,
 bookmarks=true,bookmarksnumbered=false,bookmarksopen=false,
 breaklinks=false,pdfborder={0 0 1},backref=false,colorlinks=false]
 {hyperref}
\usepackage{breakurl}

\makeatletter
 
 \@ifundefined{textcolor}{}
 {%
   \definecolor{BLACK}{gray}{0}
   \definecolor{WHITE}{gray}{1}
   \definecolor{RED}{rgb}{1,0,0}
   \definecolor{GREEN}{rgb}{0,1,0}
   \definecolor{BLUE}{rgb}{0,0,1}
   \definecolor{CYAN}{cmyk}{1,0,0,0}
   \definecolor{MAGENTA}{cmyk}{0,1,0,0}
   \definecolor{YELLOW}{cmyk}{0,0,1,0}
 }

\allowdisplaybreaks[1]

\makeatother

\begin{document}

\title{A local true Hamiltonian for the CGHS model in new variables}

\author{Saeed Rastgoo}

\affiliation{Centro de Ciencias Matemáticas, Universidad Nacional Autónoma de
México, Campus Morelia, Apartado Postal 61-3, Morelia, Michoacán 58090,
México}

\email{saeed@matmor.unam.mx}

\selectlanguage{english}%

\date{\today}
\begin{abstract}
Following our previous work, a complete classical solution of the
CGHS model in Hamiltonian formulation in new variables is given. We
preform a series of analyses and transformations to get to the CGHS
Hamiltonian in new variables from a generic class of two dimensional
dilatonic gravitational systems coupled to matter. This gives us a
second class system, a total Hamiltonian consisting of a Hamiltonian
constraint, a diffeomorphism constraint and two second class constraints.
We calculate the Dirac brackets, bring them to a standard form similar
to the Poisson brackets by introducing a new variable. Then by rescaling
lapse and shift, the Hamiltonian constraint is transformed into a
form where it has an strong Abelian algebra with itself. This property
holds both in vacuum case and in case with matter coupling. Then for
each of the vacuum and the coupled-to-matter cases, we preform two
gauge fixings, one set for each case, and solve the classical system
completely in both cases. The gauge fixing of the case coupled to
matter is done by implementing a method based on canonical transformation
to a new set of variables and leads to a local true Hamiltonian.

We also show that our formalism is consistent with the original CGHS
paper by showing that the equations of motion are the same in both
cases. Finally we derive the relevant surface term of the model.
\end{abstract}
\maketitle

\section{Introduction}

The two dimensional gravitational systems with black hole solutions,
specially the CGHS model \citep{C.G.Callan1992}, have proven to be
a very good test bench to try out various ideas about quantum gravity.
There is an extensive study of these systems in the literature (for
a short list of references see \citep{Kloesch1996,Louis-Martinez1997,Kuchar1997,Varadarajan1998,Grumiller2002,Fabbri2005,A.Ashtekar2011}
and the references within them) and there have been numerous attempts
to understand some quantum phenomena, such as black hole evaporation,
information loss and the asymptotic fate of spacetime, using their
black hole solutions. There is also the important question of whether
quantum gravity, specially loop quantum gravity, eliminates the singularity?
Loop quantum gravity has been progressing on the issue of addressing
how singularities are affected by quantum gravity. Examples are replacing
the Big Bang singularity by a Big Bounce in homogenous models \citep{Ashtekar2011}
and some form of singularity resolution for the Schwarzschild black
holes in spherically symmetric midi-superspaces \citep{Gambini2008}.
But with all these attempts, the questions surrounding these issues
has not been answered in a satisfactory way.

The purpose of this paper is twofold. On one hand, we address the
formulation of the CGHS model in a manner that is suitable for applying
loop quantum gravity techniques. We start by deriving the Ashtekar-like
variables for the model and writing its Hamiltonian in terms of those
variables. One of the differences of our method with many other works
that can be found in the literature is that this formulation is pursued
without a conformal transformation. This is important in the sense
that one is working with variables that have direct geometrical meaning
so there is no need to turn everything back to their original directly-geometric
form at the end. It also makes it easier to read the physical implications
off of the theory. Another advantage of our formulation is that it
gives us a Hamiltonian constraint that commutes with itself and hence
the algebra of the constraints becomes a Lie algebra, since now we
have structure constants instead of structure functions. Thus this
formulation of the system appears to be suitable for applying loop
quantum gravity techniques.

The second purpose of the paper is to solve the classical system coupled
to matter completely classically by some choice of gauge fixing such
that it results in a \emph{local} true Hamiltonian for the system.
Local here means that the lapse will not be an \emph{integral} of
canonical variables and hence the Hamiltonian will not be an integral
of an integral. This way the equations of motion will also be local.
The locality is important because, among other things, it is obviously
much harder to try to quantize a non-local theory. There have been
some previous studies \citep{Unruh1976,Husain2005,Gegenberg2006}
which did not lead to a Hamiltonian that was the spatial integral
of a local density, thus leading to non-local equations of motion.
This in turn leads to difficulties upon trying to quantize the system.
However, there have been some recent works \citep{Gambini2013,Alvarez2012}
leading to local true Hamiltonian for 3+1 spherically symmetric case
which we follow in order to derive such a Hamiltonian for the CGHS
model in new variables. 

The structure of this paper is as follows: in section \ref{sec:Generic-action-metric},
we rewrite a generic action for two dimensional gravitational systems
coupled to a dilaton field and a scalar matter field (which includes
the CGHS model as a special case) in a suitable way to include cases
with or without kinetic term for the dilaton field in one action.
In section \ref{sec:Tetrad-formulation}, the previous Lagrangian
will be written in tetrad variables and is transformed into a Hamiltonian
by a Legendre transformation. In section \ref{sec:The-CGHS-Hamiltonian},
we focus just on the CGHS model, derive the new Ashtekar-like variables
for it similar to the 3+1 case and write its Hamiltonian in those
variables. Sections \ref{sec:The-consistency-conditions} and \ref{sec:Dirac-bracket}
are dedicated to the implementation of the Dirac procedure for a second
class system in our Hamiltonian formalism. In section \ref{sec:Transforming-the-vacuum},
the Hamiltonian constraint is written in a special form such that
it has an strong Abelian algebra with itself. Section \ref{sec:Equations-of-motion}
is where equations of motions are derived for the Hamiltonian system.
In sections \ref{sec:Gauge-fixing-vaccum} and \ref{sec:Gauge-fixing-matter},
both vacuum and coupled-to-matter cases are gauge fixed and a local
true Hamiltonian is derived for the case coupled to matter. section
\ref{sec:Comparing-with-original} is dedicated to show that our Hamiltonian
formalism is equivalent to the original Lagrangian formalism by comparing
the equations of motion in both cases. Finally, in section \ref{sec:Boundary-terms},
the boundary term for the formalism is derived and is compared to
the standard boundary term of the CGHS model.

\section{Generic action in metric variables\label{sec:Generic-action-metric}}

It has been shown \citep{Banks1991,Odintsov1991,Kloesch1996} that
the most general diffeomorphism invariant action yielding second order
differential equations for the metric $g$ and a scalar (dilaton)
field $\Phi$ in two dimensions is%
\footnote{The authors in \citep{Grumiller2002} argue that the most general
form is in fact $S=\int d^{2}x\sqrt{-|g|}\left(D(\Phi)R(g)+V\left(\left(\nabla\Phi\right)^{2},\Phi\right)\right)$
.%
}
\begin{equation}
S_{g\textnormal{-dil}}=\int d^{2}x\sqrt{-|g|}\left(D(\Phi)R(g)+\frac{1}{2}g^{ab}\partial_{a}\Phi\partial_{b}\Phi+V(\Phi)\right).\label{eq:dil-full-g}
\end{equation}
In fact two of the most important models of gravitational systems
are examples of this action. The first case, the 4D spherically symmetric
model, is just the Einstein-Hilbert action. Minimally coupled to matter
it is
\begin{equation}
S_{\textrm{sph}}=\frac{1}{16\pi}\int d^{4}x\sqrt{-|\bar{g}|}R-\frac{1}{4\pi}\int d^{4}x\sqrt{-|\bar{g}|}\bar{g}^{ab}\partial_{a}f\partial_{b}f,
\end{equation}
which by using spherically symmetric ansatz
\begin{align}
ds^{2}= & g_{\mu\nu}dx^{\mu}dx^{\nu}+\Phi^{2}(d\theta^{2}+\sin^{2}(\theta)d\phi^{2}), & \ensuremath{\mu,\nu=0,1}\label{eq:sph-ansatz}
\end{align}
and integrating over $\theta$ and $\phi$, becomes
\begin{equation}
S_{\textrm{spher}}=\int d^{2}x\sqrt{-|g|}\left(\frac{1}{4}\Phi^{2}R(g)+\frac{1}{2}g^{ab}\partial_{a}\Phi\partial_{b}\Phi+\frac{1}{2}\right)-\frac{1}{2}\int d^{2}x\sqrt{-|g|}\Phi^{2}g^{ab}\partial_{a}f\partial_{b}f.\label{eq:spher-dil}
\end{equation}
In (\ref{eq:sph-ansatz}), $x^{0}$, $x^{1}$, $\theta$ and $\phi$
are some coordinates adapted to the spherical symmetry and $g_{\mu\nu}$
is the metric on the $x^{0},x^{1}$ plane. If $\frac{\partial\Phi}{\partial x^{1}}\neq0$,
it is possible to rescale $\Phi$ to be identical to $x^{1}$. It
is easily seen that (the gravitational part of) (\ref{eq:spher-dil})
is an example of (\ref{eq:dil-full-g}) and for this case, $\Phi$
mimics the dilation field.

The second case is the CGHS model \citep{C.G.Callan1992} whose action
minimally coupled to matter is
\begin{equation}
S_{\textrm{CGHS}}=\int d^{2}x\sqrt{-|g|}e^{-2\varphi}\left(R+4g^{ab}\partial_{a}\varphi\partial_{b}\varphi+4\lambda^{2}\right)-\frac{1}{2}\int d^{2}x\sqrt{-|g|}g^{ab}\partial_{a}f\partial_{b}f,
\end{equation}
in which $\varphi$ (not to be confused with the coordinate $\phi$
in spherically symmetric model) corresponds to the dilaton field and
$\lambda^{2}$ is the cosmological constant. By a redefinition of
the dilation field
\begin{equation}
\Phi=2\sqrt{2}e^{-\varphi},
\end{equation}
one gets
\begin{equation}
S_{\textrm{CGHS}}=\int d^{2}x\sqrt{-|g}|\left\{ \frac{1}{8}\Phi^{2}R+\frac{1}{2}g^{ab}\partial_{a}\Phi\partial_{b}\Phi+\frac{1}{2}\Phi^{2}\lambda^{2}\right\} -\frac{1}{2}\int d^{2}x\sqrt{-|g|}g^{ab}\partial_{a}f\partial_{b}f
\end{equation}
whose gravitation part can be seen to be an example of (\ref{eq:dil-full-g}).
So from these two cases we can infer that the minimally coupled matter
part corresponding to (\ref{eq:dil-full-g}), can be written as
\begin{equation}
S_{\textrm{m}}=-\int d^{2}x\sqrt{-|g|}W(\Phi)g^{ab}\partial_{a}f\partial_{b}f.
\end{equation}
Thus the full general action with minimal coupling to matter will
become
\begin{equation}
S=\int d^{2}x\sqrt{-|g|}\left(D(\Phi)R(g)+\frac{1}{2}g^{ab}\partial_{a}\Phi\partial_{b}\Phi+V(\Phi)\right)-\int d^{2}x\sqrt{-|g|}W(\Phi)g^{ab}\partial_{a}f\partial_{b}f.\label{eq:dil-2}
\end{equation}
In some cases, it is desirable to eliminate the kinetic term $g^{ab}\partial_{a}\Phi\partial_{b}\Phi$.
This can be achieved by a conformal transformation
\begin{equation}
\tilde{g}_{ab}=\Omega^{2}(\Phi)g_{ab},
\end{equation}
with
\begin{equation}
\Omega(\Phi)=C\exp\left(\frac{1}{4}\int d\Phi\frac{1}{\frac{dD(\Phi)}{d\Phi}}\right)
\end{equation}
and $C$ being a constant of integration. In this case the general
action becomes
\begin{equation}
S=\int d^{2}x\sqrt{-|\tilde{g}|}\left(D(\Phi)\tilde{R}(g)+\Omega^{-2}(\Phi)V(\Phi)\right)-\int d^{2}x\sqrt{-|\tilde{g}|}W(\Phi)\tilde{g}^{ab}\partial_{a}f\partial_{b}f.\label{eq:dil-3}
\end{equation}
In order to keep the generality, we can combine (\ref{eq:dil-2})
and (\ref{eq:dil-3}) into the following form
\begin{equation}
S_{\textrm{1+1}}=\int d^{2}x\sqrt{-|g|}\left\{ Y(\Phi)R+\frac{1}{2}Zg^{ab}\partial_{a}\Phi\partial_{b}\Phi+V(\Phi)\right\} -\int d^{2}x\sqrt{-|g|}W(\Phi)g^{ab}\partial_{a}f\partial_{b}f,\label{eq:Generic-Action}
\end{equation}
where we have introduced the variable $Z=0,1$, which plays the rule
of a ``switch'' that turns the kinetic term on or off. The gravitational
and matter Lagrangian densities are now
\begin{align}
L_{\textnormal{g}} & =\sqrt{-|g|}\left\{ Y(\Phi)R+\frac{1}{2}Zg^{ab}\partial_{a}\Phi\partial_{b}\Phi+V(\Phi)\right\} ,\label{eq:Generic-Lag-g}\\
L_{\textnormal{m}} & =-\sqrt{-|g|}W(\Phi)g^{ab}\partial_{a}f\partial_{b}f.\label{eq:Generic-Lag-m}
\end{align}
It is worth noting that the action (\ref{eq:Generic-Action}) is similar
to the action of $f(R)$ gravity theories with $V(\Phi)$ being the
potential of the dilaton field. In the CGHS case, this potential is
related to the cosmological constant. So the cosmological constant
in this case may be seen as dynamical and coming from properties of
the dilaton field rather than just being there.

\section{Tetrad formulation\label{sec:Tetrad-formulation}}

In order to get to the new variables for these cases, one needs to
first write the theory in tetrad formulation in which
\begin{equation}
g_{ab}=\eta_{IJ}e^{I}{}_{a}e^{J}{}_{b}
\end{equation}
and $\eta_{IJ}$ is the Minkowski metric, $e^{I}{}_{a}$ are the tetrads
and $I,J$ are the internal indices while $a,b$ are the abstract
ones. The curvature can be written in terms of the curvature of the
spin connection and ultimately in terms of the spin connection $\omega_{a}{}^{IJ}$
itself as 
\begin{align}
R & =R_{ab}{}^{IJ}e^{a}{}_{I}e^{b}{}_{J}\nonumber \\
 & =\left(2\partial_{[a}\omega_{a]}{}^{IJ}+[\omega_{a},\omega_{b}]^{IJ}\right)e^{a}{}_{I}e^{b}{}_{J},\label{eq:curv-sp-con}
\end{align}
where $[,]$ stands for the Lie commutator in the Lorentz Lie algebra
and the indices $I,J$ take value in this algebra. Since the spin
connection is antisymmetric in $I,J$, we can write it as
\begin{equation}
\omega_{a}{}^{IJ}=\omega_{a}\epsilon^{IJ}.
\end{equation}
This way, the curvature (\ref{eq:curv-sp-con}) becomes
\begin{align}
R & =\left(2\partial_{[a}\omega_{a]}\epsilon^{IJ}+\omega_{[a}\omega_{b]}\eta^{IJ}\right)e^{a}{}_{I}e^{b}{}_{J}\nonumber \\
 & =2\partial_{[a}\omega_{a]}\epsilon^{IJ}e^{a}{}_{I}e^{b}{}_{J},
\end{align}
where we have used the following fact about the Lie commutator in
this case
\begin{align}
[\omega_{a},\omega_{b}]^{IJ} & =\omega_{a}\epsilon^{I}{}_{K}\omega_{b}\epsilon^{KJ}-\omega_{b}\epsilon^{I}{}_{K}\omega_{a}\epsilon^{KJ}\nonumber \\
 & =\omega_{[a}\omega_{b]}\underbrace{\epsilon^{I}{}_{K}\epsilon^{KJ}}_{\eta^{IJ}}
\end{align}
and also the fact that $\eta^{IJ}e^{a}{}_{I}e^{b}{}_{J}$ is symmetric
in $a,b$ while $\omega_{[a}\omega_{b]}$ is antisymmetric and thus
\begin{equation}
\omega_{[a}\omega_{b]}\eta^{IJ}e^{a}{}_{I}e^{b}{}_{J}=0.
\end{equation}
We also would like to add the torsion free condition (contracted by
a Lagrange multiplier) to the action (\ref{eq:Generic-Action}). This
condition reads 
\begin{equation}
0=de^{I}+\epsilon^{I}{}_{J}\omega\wedge e^{J}=2\partial_{[a}e_{b]}{}^{I}+2\epsilon^{I}{}_{J}\omega_{[a}e_{b]}{}^{J}.
\end{equation}
Since this is a 2-form, we need to contract it with $\epsilon^{ab}$
to get an scalar to be able to add it to the Lagrangian density. Doing
so, contracting it with a Lagrange multiplier $-X_{I}$ and substituting
everything in (\ref{eq:Generic-Lag-g}) yields
\begin{equation}
L_{\textnormal{g}}=-2X_{I}\epsilon^{ab}(\partial_{[a}e_{b]}{}^{I}+\epsilon^{I}{}_{J}\omega_{[a}e_{b]}{}^{J})+2Y\partial_{[a}\omega_{b]}e\epsilon^{IJ}e^{a}{}_{I}e^{b}{}_{J}+\frac{1}{2}Z\eta^{IJ}ee_{I}{}^{a}e_{J}{}^{b}\partial_{a}\Phi\partial_{b}\Phi+eV.\label{eq:act-tet1}
\end{equation}
We can write $\epsilon^{ab}$ in terms of $\epsilon^{IJ},\ e_{a}{}^{I}$
and its determinant $e=\det(e_{a}{}^{I})$ as
\begin{equation}
\epsilon^{ab}=-ee_{I}{}^{a}e_{J}{}^{b}\epsilon^{IJ}.
\end{equation}
If we integrate by parts in the first term in Lagrangian density (\ref{eq:act-tet1})
to bring the partial derivative to act on $X_{I}$, and then use the
above result for $\epsilon^{ab}$ in both first two terms of this
Lagrangian, the pure gravitational Lagrangian density can be written
as 
\begin{align}
L_{\textrm{g}}= & e\left(-2\partial_{a}(X_{I})e_{K}{}^{a}\epsilon^{KI}-2X_{I}e^{Ia}\omega_{a}+2Y\partial_{a}\omega_{b}\epsilon^{IJ}e_{I}{}^{a}e_{J}{}^{b}+\frac{1}{2}Z\eta^{IJ}e_{I}{}^{a}e_{J}{}^{b}\partial_{a}\Phi\partial_{b}\Phi+V\right).\label{eq:act-tet-ger}
\end{align}
From (\ref{eq:Generic-Lag-m}), the matter Lagrangian density can
also simply be written as 
\begin{equation}
L_{\textnormal{m}}=-W\eta^{IJ}ee_{I}{}^{a}e_{J}{}^{b}\partial_{a}f\partial_{b}f.\label{eq:act-tet-m}
\end{equation}
Next, we decompose the Lagrangian by ADM method and perform a Legendre
transformation to get to the Hamiltonian. Most of the details needed
for these steps have been already discussed in our previous work \citep{us-new-var}.
So we just mention the result here. The generic Hamiltonian then will
become.
\begin{align}
H= & N\Bigg[2\frac{P_{2}}{|P|}\left(^{*}X^{1}\right)'+2\frac{P_{1}}{|P|}\left(^{*}X^{2}\right)'-2\frac{P_{1}}{|P|}\omega_{1}{}^{*}X^{1}-2\frac{P_{2}}{|P|}\omega_{1}{}^{*}X^{2}\nonumber \\
 & -\frac{Z}{|P|}\Phi^{\prime2}-\frac{P_{\Phi}^{2}}{Z|P|}+\frac{2Wf^{\prime2}}{|P|}+\frac{P_{f}^{2}}{2W|P|}-\frac{|P|}{2}V\Bigg]\nonumber \\
 & +N^{1}\left[P_{\Phi}\Phi'+P_{f}f'+P_{1}\left(^{*}X^{1}\right)'+P_{2}\left(^{*}X^{2}\right)'-P_{1}\omega_{1}{}^{*}X^{2}-P_{2}\omega_{1}{}^{*}X^{1}\right]\nonumber \\
 & +\omega_{0}\left[P_{1}{}^{*}X^{2}+P_{2}{}^{*}X^{1}-\left(2Y\right)'\right].\label{eq:H3}
\end{align}
Here $N$ is lapse, $N^{1}$ is the (one dimensional) shift vector,
the momenta are
\begin{align}
P_{I}= & \frac{\partial L}{\partial{}^{*}\dot{X}^{I}}=2\sqrt{q}n_{I},\label{eq:PI1}\\
P_{\omega}= & \frac{\partial L}{\partial\dot{\omega}_{1}}=2Y,\label{eq:Pomega1}\\
P_{\Phi}= & \frac{\partial L}{\partial\dot{\Phi}}=\frac{Z\sqrt{q}}{N}\left(N^{1}\Phi'-\dot{\Phi}\right),\label{eq:Pphi1}\\
P_{f}= & \frac{\partial L}{\partial\dot{f}}=-\frac{2W\sqrt{q}}{N}\left(N^{1}f'-\dot{f}\right),\label{eq:Pf1}
\end{align}
$n_{I}$'s are normals to the spatial hypersurfaces, $^{*}X^{I}=\epsilon^{IJ}X_{J}$,
$q$ is the determinant of the induced metric on the spatial hypersurfaces,
$|P|=\sqrt{-\eta^{IJ}P_{I}P_{J}}$ is the norm of $P_{I}$ (which
is a timelike vector, hence the negative sign) and the prime represents
partial derivative with respect to the spatial coordinate $x^{1}$.
From the definitions of momenta (\ref{eq:Pomega1}) and (\ref{eq:Pphi1}),
one can see that if $Z=1$, i.e. cases with kinetic term present (without
conformal transformation), then $\Phi$ is a canonical variable and
$P_{\Phi}$ is its momentum %
\footnote{In fact this can be seen more clearly in the Lagrangian which leads
to this Hamiltonian. There, all the terms containing $\dot{\Phi}$
are multiplied by $Z$.%
}. In this case, since $Y$ involves $\Phi$, equation (\ref{eq:Pomega1})
will be a new primary constraint. This would not happen in the cases
without kinetic term (with conformal transformation) like standard
spherically symmetric case in Ashtekar variables, since there, $\Phi$
is not a canonical variable although equation (\ref{eq:Pomega1})
is still valid. Thus in the CGHS case, we should add (\ref{eq:Pomega1})
to the general Hamiltonian (\ref{eq:H3}) to get the total Hamiltonian.

\section{The CGHS Hamiltonian in new variables\label{sec:The-CGHS-Hamiltonian}}

From now on we focus only on the CGHS case and hence we substitute
the explicit forms of $Y(\Phi),\, V(\Phi)$ and $W(\Phi)$ in the
previous generic Hamiltonian and also add (\ref{eq:Pomega1}) as a
new primary constraint to it to get
\begin{align}
H= & N\Bigg(\frac{2P_{2}}{|P|}\partial_{1}{}^{*}X^{1}+\frac{2P_{1}}{|P|}\partial_{1}{}^{*}X^{2}-\frac{2P_{1}}{|P|}\omega_{1}{}^{*}X^{1}-\frac{2P_{2}}{|P|}\omega_{1}{}^{*}X^{2}-\frac{|P|}{4}\lambda^{2}\Phi^{2}\nonumber \\
 & -\frac{\Phi^{\prime2}}{|P|}-\frac{P_{\Phi}^{2}}{|P|}+\frac{(f')^{2}}{|P|}+\frac{P_{f}^{2}}{|P|}\Bigg)\nonumber \\
 & +N^{1}\left(P_{1}\partial_{1}{}^{*}X^{1}+P_{2}\partial_{1}{}^{*}X^{2}-P_{2}{}^{*}X^{1}\omega_{1}-P_{1}{}^{*}X^{2}\omega_{1}+\Phi'P_{\Phi}+f'P_{f}\right)\nonumber \\
 & +\omega_{0}\left(P_{1}{}^{*}X^{2}+P_{2}{}^{*}X^{1}-\left(\frac{1}{4}\Phi^{2}\right)'\right)\nonumber \\
 & +M\left(P_{\omega}-\frac{1}{4}\Phi^{2}\right),\label{eq:Htot-preAsh}
\end{align}
where $M$ is a Lagrange multiplier. In order to transform to the
Ashtekar variables and following a similar pattern as for Ashtekar
variables in the 3+1 model \citep{us-new-var}, we introduce the following
new momenta with a canonical transformation: 
\begin{align}
P_{\omega}= & E^{x},\label{eq:Pomega-Ash}\\
|P|= & 2E^{\varphi},\label{eq:Pnorm-Ash}\\
P_{1}= & 2\cosh(\eta)E^{\varphi},\label{eq:P1-Ash}\\
P_{2}= & 2\sinh(\eta)E^{\varphi},\label{eq:P2-Ash}
\end{align}
where $E^{x},\, E^{\varphi}$ and $\eta$ are the new momenta. This
gives us the generating function 
\begin{equation}
F(q,P)=2{}^{*}X^{1}\cosh(\eta)E^{\varphi}+2{}^{*}X^{2}\sinh(\eta)E^{\varphi}+\omega_{1}E^{x}+\Phi P_{\Phi}+fP_{f}.
\end{equation}
Using $F(q,P)$, we can find the new canonical variables as 
\begin{align}
Q_{\eta}= & \frac{\partial F}{\partial\eta}=2{}^{*}X^{1}\sinh(\eta)E^{\varphi}+2{}^{*}X^{2}\cosh(\eta)E^{\varphi},\\
K_{\varphi}= & \frac{\partial F}{\partial E^{\varphi}}=2{}^{*}X^{1}\cosh(\eta)+2{}^{*}X^{2}\sinh(\eta),\\
A_{x}= & \frac{\partial F}{\partial E^{x}}=\omega_{1},
\end{align}
where $Q_{\eta},\, K_{\varphi}$ and $A_{x}$ correspond to $\eta,\, E^{\varphi}$
and $E^{x}$ respectively. From the above equations, we can find $^{*}X^{1},\ {}^{*}X^{2}$
and $\omega_{1}$ as 
\begin{align}
^{*}X^{1}= & \frac{1}{2}\left(K_{\varphi}\cosh(\eta)-\frac{Q_{\eta}\sinh(\eta)}{E^{\varphi}}\right),\label{eq:X1-Ash}\\
^{*}X^{2}= & -\frac{1}{2}\left(K_{\varphi}\sinh(\eta)-\frac{Q_{\eta}\cosh(\eta)}{E^{\varphi}}\right),\label{eq:X2-Ash}\\
\omega_{1}= & A_{x}.\label{eq:omega1-Ash}
\end{align}
In order to write the Hamiltonian density (\ref{eq:Htot-preAsh})
in these new variables, we substitute (\ref{eq:Pomega-Ash})-(\ref{eq:P2-Ash})
and (\ref{eq:X1-Ash})-(\ref{eq:omega1-Ash}) in the total Hamiltonian
(\ref{eq:Htot-preAsh}), and then make a field redefinition 
\begin{equation}
A_{x}=K_{x}-\eta'
\end{equation}
to get rid of $\eta$ in the Hamiltonian and get 
\begin{align}
H= & N\left(\frac{Q_{\eta}'}{E^{\varphi}}-\frac{Q_{\eta}E^{\varphi\prime}}{E^{\varphi2}}-\frac{1}{2}E^{\varphi}\lambda^{2}\Phi^{2}-K_{\varphi}K_{x}-\frac{\Phi^{\prime2}}{2E^{\varphi}}-\frac{P_{\Phi}^{2}}{2E^{\varphi}}+\frac{(f')^{2}}{2E^{\varphi}}+\frac{P_{f}^{2}}{2E^{\varphi}}\right)\nonumber \\
 & +N^{1}\left(E^{\varphi}K_{\varphi}'-Q_{\eta}K_{x}+\Phi'P_{\Phi}+f'P_{f}\right)\nonumber \\
 & +\omega_{0}\left(Q_{\eta}-\left(\frac{1}{4}\Phi^{2}\right)'\right)+M\left(E^{x}-\frac{1}{4}\Phi^{2}\right).
\end{align}
We can see from here that the total Hamiltonian is just the sum of
four constraints as is expected for a totally constrained system.
The first constraint multiplied by the lapse function $N$ is the
Hamiltonian constraint. The one that is multiplied by the shift vector
$N^{1}$ is the diffeomorphism constraint. The constraint that is
multiplied by $\omega_{0}$ is the Gauss constraint and the last one
is the one we got from the definition of the momentum $P_{\omega}$.
Solving the Gauss constraint in the above Hamiltonian and substituting
the resultant $Q_{\eta}$ from it back into the Hamiltonian yields

\begin{align}
H= & N\left(\frac{\Phi\Phi''}{2E^{\varphi}}-\frac{\Phi\Phi'E^{\varphi\prime}}{2E^{\varphi2}}-\frac{1}{2}E^{\varphi}\lambda^{2}\Phi^{2}-K_{\varphi}K_{x}-\frac{P_{\Phi}^{2}}{2E^{\varphi}}+\frac{(f')^{2}}{2E^{\varphi}}+\frac{P_{f}^{2}}{2E^{\varphi}}\right)\nonumber \\
 & +N^{1}\left(E^{\varphi}K_{\varphi}'-\frac{1}{2}\Phi\Phi'K_{x}+\Phi'P_{\Phi}+f'P_{f}\right)+M\left(E^{x}-\frac{1}{4}\Phi^{2}\right).\label{eq:Htot-preAsh-2}
\end{align}
Since we now know our canonical variables and momenta, we can write
their Poisson brackets as
\begin{align}
\{K_{x}(x),E^{x}(y)\} & =\delta(x-y),\label{eq:pois1-1}\\
\{K_{\varphi}(x),E^{\varphi}(y)\} & =\delta(x-y),\label{eq:pois1-2}\\
\{\Phi(x),P_{\Phi}(y)\} & =\delta(x-y),\label{eq:pois1-3}\\
\{f(x),P_{f}(y)\} & =\delta(x-y),\label{eq:pois1-4}
\end{align}
and we have not written the Poisson bracket of $(Q_{\eta},\eta)$
pair because they no longer appear in the Hamiltonian. The rest of
the Poisson brackets are strongly zero.

\section{The consistency conditions on constraints\label{sec:The-consistency-conditions}}

Following the Dirac procedure, we should check the preservation of
the constraints to see if there are any new secondary constraints
and/or to find the value of the Lagrange multipliers in terms of canonical
variables. This means that the constraints $\mathcal{C}$, being the
constants of motion, should remain weakly vanishing during the evolution
\begin{equation}
\dot{\mathcal{C}}=\{\mathcal{C},H\}\approx0,
\end{equation}
where $\approx$ represents weak inequality. The Poisson brackets
of the Hamiltonian and diffeomorphism constraints with $H$ vanishes
weakly. Let's check the consistency of the constraint 
\begin{equation}
{\displaystyle \mu=E^{x}-\frac{1}{4}\Phi^{2}}.\label{eq:muconst}
\end{equation}
For this, we need (\ref{eq:pois1-1}) and (\ref{eq:pois1-3}). The
preservation condition of $\mu$ constraint leads to a new, and by
definition secondary, constraint which we call $\alpha$: 
\begin{equation}
\dot{\mu}=\{\mu,H\}\approx0\Rightarrow\alpha=K_{\varphi}+\frac{1}{2}\frac{P_{\Phi}\Phi}{E^{\varphi}}\approx0.\label{eq:alphaconst}
\end{equation}
We also need to check the preservation of the new $\alpha$ constraint.
This leads to a relation between the Lagrange multipliers $N,\ N^{1}$
and $M$ (and canonical variables). Finding $M$ from this relation
and substituting it into the total Hamiltonian (\ref{eq:Htot-preAsh-2})
yields 
\begin{align}
H=N\bigg( & -K_{\varphi}K_{x}-\frac{2\Phi'E^{\varphi\prime}E^{x}}{\Phi E^{\varphi2}}+\frac{2E^{x}\Phi''}{\Phi E^{\varphi}}-\frac{P_{\Phi}^{2}}{2E^{\varphi}}-\frac{1}{2}E^{\varphi}\lambda^{2}\Phi^{2}-\frac{1}{2}\frac{\Phi P_{\Phi}K_{x}}{E^{\varphi}}+\frac{2P_{\Phi}K_{x}E^{x}}{\Phi E^{\varphi}}\nonumber \\
 & +\frac{2E^{x}P_{f}^{2}}{\Phi^{2}E^{\varphi}}-\frac{2\Phi^{\prime2}E^{x}}{\Phi^{2}E^{\varphi}}+\frac{\Phi^{\prime2}}{2E^{\varphi}}+\frac{2E^{x}f^{\prime2}}{\Phi^{2}E^{\varphi}}\bigg)\nonumber \\
+N^{1}\bigg( & -\frac{1}{2}\Phi\Phi'K_{x}+\Phi'P_{\Phi}+f'P_{f}+E^{x}K_{x}'+E^{\varphi}K_{\varphi}'-\frac{1}{4}\Phi^{2}K_{x}'\bigg).\label{eq:Htot-cons}
\end{align}
Next step is to check if the constraints are first class or second
class. Calculating the Poisson brackets of constraints among themselves
shows that $\mu$ and $\alpha$ are second class and do not commute
with each other. In other words, their Poisson bracket with each other
does not vanish weakly. Since there are second class constraints in
the theory, now we should abandon the Poisson bracket and move on
to the Dirac bracket and also put the second class constraint strongly
equal to zero and eliminate some of the variables in term of others.
By doing this, we can get rid of the $(\Phi,P_{\Phi})$ pair in the
Hamiltonian as can be seen below. Equating both the $\mu$ and $\alpha$
constraints strongly to zero yields 
\begin{align}
\mu=0 & \Rightarrow\Phi=2\sqrt{E^{x}},\\
\alpha=0 & \Rightarrow P_{\Phi}=-\frac{K_{\varphi}E^{\varphi}}{\sqrt{E^{x}}}.
\end{align}
Substituting these in the total Hamiltonian (\ref{eq:Htot-cons})
yields 
\begin{align}
H= & N\left(-K_{\varphi}K_{x}-\frac{E^{\varphi\prime}E^{x\prime}}{E^{\varphi2}}-\frac{1}{2}\frac{E^{x\prime2}}{E^{\varphi}E^{x}}+\frac{E^{x\prime\prime}}{E^{\varphi}}-\frac{1}{2}\frac{K_{\varphi}^{2}E^{\varphi}}{E^{x}}-2E^{\varphi}E^{x}\lambda^{2}+\frac{1}{2}\frac{P_{f}^{2}}{E^{\varphi}}+\frac{1}{2}\frac{f^{\prime2}}{E^{\varphi}}\right)\nonumber \\
 & +N^{1}\left(-K_{x}E^{x\prime}+f'P_{f}-\frac{K_{\varphi}E^{\varphi}E^{x\prime}}{E^{x}}+E^{\varphi}K_{\varphi}'\right).\label{eq:Htot-fin}
\end{align}
where now we are only left with a Hamiltonian and a diffeomorphism
constraint.

\section{Dirac bracket and the algebra of canonical variables\label{sec:Dirac-bracket}}

In order to switch to the Dirac bracket, we need to find the general
form of the Dirac bracket for our theory. For a field theory (where
the variables have continuous indices), the Dirac bracket is
\begin{equation}
\{A(x),B(y)\}_{D}=\{A(x),B(y)\}-\int dw\int dz\left(\{A(x),\chi_{\rho}(w)\}C^{\rho\sigma}(w,z)\{\chi_{\sigma}(z),B(y)\}\right),\label{eq:Diracgeneral}
\end{equation}
where the $\{,\}_{D}$ refers to the Dirac bracket, $\chi$'s are
the second class constraints and $C^{\rho\sigma}(w,z)$ are the elements
of the inverse of the matrix of the Poisson brackets between the $\rho$'th
and $\sigma$'th second class constraints
\begin{equation}
C_{\rho\sigma}(w,z)=\{\chi_{\rho}(w),\chi_{\sigma}(z)\}.
\end{equation}
In our model, there are only two second class constraints, $\mu$
and $\alpha$. Thus the matrix of the Poisson brackets of the second
class constraints will be
\begin{align}
\mathbf{C}=C_{\rho\sigma}(x,y)= & \begin{pmatrix}\{\mu(x),\mu(y)\} & \{\mu(x),\alpha(y)\}\\
\{\alpha(x),\mu(y)\} & \{\alpha(x),\alpha(y)\}
\end{pmatrix}\nonumber \\
= & \begin{pmatrix}0 & \{\mu(x),\alpha(y)\}\\
\{\alpha(x),\mu(y)\} & 0
\end{pmatrix}.
\end{align}
To compute the elements of this matrix we use (\ref{eq:muconst})
and (\ref{eq:alphaconst}) along with (\ref{eq:pois1-3}) to get
\begin{align}
\{\mu(x),\alpha(y)\} & =\left\{ E^{x}(x)-\frac{1}{4}\Phi(x)^{2},K_{\varphi}(y)+\frac{1}{2}\frac{P_{\Phi}(y)\Phi(y)}{E^{\varphi}(y)}\right\} \nonumber \\
 & =-\frac{1}{8}\frac{\Phi(y)}{E^{\varphi}(y)}\left\{ \Phi(x)^{2},P_{\Phi}(y)\right\} \nonumber \\
 & =-\frac{1}{4}\frac{\Phi(y)^{2}}{E^{\varphi}(y)}\delta(x-y).
\end{align}
The same method of computations gives
\begin{equation}
\{\mu(x),\alpha(y)\}=\frac{1}{4}\frac{\Phi(x)^{2}}{E^{\varphi}(x)}\delta(x-y).
\end{equation}
To calculate the elements of $\mathbf{C}^{-1}$, we use the property
$\mathbf{C}\mathbf{C}^{-1}=\mathbf{1}$, or in terms of their elements
\begin{equation}
\int C_{\rho\sigma}(x,z)C^{\sigma\beta}(z,y)dz=\delta_{\rho}{}^{\beta}\delta(x-y),
\end{equation}
which yields
\begin{equation}
\mathbf{C}^{-1}=C^{\rho\sigma}(x,y)=\begin{pmatrix}0 & \frac{4E^{\varphi}(x)}{\Phi^{2}(x)}\\
-\frac{4E^{\varphi}(x)}{\Phi^{2}(x)} & 0
\end{pmatrix}\delta(x-y).
\end{equation}
Using this and (\ref{eq:Diracgeneral}), the general form of the Dirac
bracket for our theory becomes 
\begin{align}
\{A(x),B(y)\}_{D}= & \{A(x),B(y)\}+\int dw\int dz\left(\{A(x),\mu(w)\}\frac{4E^{\varphi}(w)}{\Phi^{2}(w)}\delta(w-z)\{\alpha(z),B(y)\}\right)\nonumber \\
 & -\int dw\int dz\left(\{A(x),\alpha(w)\}\frac{4E^{\varphi}(w)}{\Phi^{2}(w)}\delta(w-z)\{\mu(z),B(y)\}\right).\label{eq:Dirac-general}
\end{align}
If we use this formula and the Poisson brackets (\ref{eq:pois1-1})-(\ref{eq:pois1-4}),
we can find the Dirac brackets of the canonical variables between
each other as 
\begin{align}
\{K_{x}(x),E^{x}(y)\}_{D} & =\{K_{\varphi}(x),E^{\varphi}(y)\}_{D}=\{f(x),P_{f}(y)\}_{D}=\delta(x-y),\label{eq:Dirac-Br-pairs}\\
\{K_{x}(x),K_{\varphi}(y)\}_{D} & ={\displaystyle \frac{K_{\varphi}}{E^{x}}\delta(x-y)},\label{eq:Dirac-Br-KxKf}\\
\{K_{x},E^{\varphi}\}_{D} & =-{\displaystyle \frac{E^{\varphi}}{E^{x}}\delta(x-y)},\label{eq:Dirac-Br-KxEf}\\
\{E^{x},K_{\varphi}\}_{D} & =\{E^{x},E^{\varphi}\}_{D}=\{f,\blacksquare\}_{D}=\{P_{f},\bullet\}_{D}=0,\label{eq:Dirac-Br-other}
\end{align}
where $\blacksquare$ means everything except $P_{f}$ and the $\bullet$
means everything except $f$.

Looking at the above Dirac brackets, we can make an important observation:
by introducing a new variable
\begin{equation}
U_{x}=K_{x}+\frac{E^{\varphi}K_{\varphi}}{E^{x}},\label{eq:Ux-Def}
\end{equation}
the Dirac brackets (\ref{eq:Dirac-Br-pairs})-(\ref{eq:Dirac-Br-other})
can be brought to the standard from
\begin{equation}
\{U_{x}(x),E^{x}(y)\}_{D}=\{K_{\varphi}(x),E^{\varphi}(y)\}_{D}=\{f(x),P_{f}(y)\}_{D}=\delta(x-y),\label{eq:brackets-main}
\end{equation}
with other brackets being zero. This is an important step since it
makes Dirac brackets look like the standard form of the Poisson brackets
of canonical pairs. Using (\ref{eq:Ux-Def}), the total Hamiltonian
(\ref{eq:Htot-fin}) becomes
\begin{align}
H= & N\left(-K_{\varphi}U_{x}-\frac{E^{\varphi\prime}E^{x\prime}}{E^{\varphi2}}-\frac{1}{2}\frac{E^{x\prime2}}{E^{\varphi}E^{x}}+\frac{E^{x\prime\prime}}{E^{\varphi}}+\frac{1}{2}\frac{K_{\varphi}^{2}E^{\varphi}}{E^{x}}-2E^{\varphi}E^{x}\lambda^{2}+\frac{1}{2}\frac{P_{f}^{2}}{E^{\varphi}}+\frac{1}{2}\frac{f^{\prime2}}{E^{\varphi}}\right)\nonumber \\
 & +N^{1}\left(-U_{x}E^{x\prime}+f'P_{f}+E^{\varphi}K_{\varphi}'\right).\label{eq:HtotfnUx}
\end{align}

\section{Transforming the vacuum Hamiltonian constraint into a total derivative\label{sec:Transforming-the-vacuum}}

We can omit $U_{x}$ in the Hamiltonian constraint in (\ref{eq:HtotfnUx})
by a redefinition of the shift
\begin{equation}
\overline{N}^{1}=N^{1}+\frac{NK_{\varphi}}{E^{x\prime}}.\label{eq:N1-redef}
\end{equation}
Substituting this into (\ref{eq:HtotfnUx}) gives
\begin{align}
H= & N\bigg(-\frac{E^{\varphi\prime}E^{x\prime}}{E^{\varphi2}}-\frac{1}{2}\frac{E^{x\prime2}}{E^{\varphi}E^{x}}+\frac{E^{x\prime\prime}}{E^{\varphi}}+\frac{1}{2}\frac{K_{\varphi}^{2}E^{\varphi}}{E^{x}}-\frac{E^{\varphi}K_{\varphi}K_{\varphi}^{\prime}}{E^{x\prime}}\nonumber \\
 & -\frac{f^{\prime}P_{f}K_{\varphi}}{E^{x\prime}}-2E^{\varphi}E^{x}\lambda^{2}+\frac{1}{2}\frac{P_{f}^{2}}{E^{\varphi}}+\frac{1}{2}\frac{f^{\prime2}}{E^{\varphi}}\bigg)\nonumber \\
 & +\overline{N}^{1}\left(-U_{x}E^{x\prime}+f'P_{f}+E^{\varphi}K_{\varphi}'\right).
\end{align}
Now redefining the lapse in the above total Hamiltonian as
\begin{equation}
\overline{N}=N\frac{E^{\varphi}E^{x}}{E^{x\prime}}\label{eq:N-redef}
\end{equation}
will yield
\begin{align}
H_{T}= & \overline{N}\left(\frac{\partial}{\partial x}\left(\frac{1}{2}\frac{E^{x\prime2}}{E^{\varphi2}E^{x}}-2E^{x}\lambda^{2}-\frac{1}{2}\frac{K_{\varphi}^{2}}{E^{x}}\right)-\frac{f^{\prime}P_{f}K_{\varphi}}{E^{x}E^{\varphi}}+\frac{1}{2}\frac{P_{f}^{2}E^{x\prime}}{E^{\varphi2}E^{x}}+\frac{1}{2}\frac{E^{x\prime}f^{\prime2}}{E^{\varphi2}E^{x}}\right)\nonumber \\
 & +\overline{N}^{1}\left(-U_{x}E^{x\prime}+f'P_{f}+E^{\varphi}K_{\varphi}'\right).\label{eq:HT-deriv-mat}
\end{align}
The Hamiltonian constraint above
\begin{equation}
\mathcal{H}=\frac{\partial}{\partial x}\left(\frac{1}{2}\frac{E^{x\prime2}}{E^{\varphi2}E^{x}}-2E^{x}\lambda^{2}-\frac{1}{2}\frac{K_{\varphi}^{2}}{E^{x}}\right)-\frac{f^{\prime}P_{f}K_{\varphi}}{E^{x}E^{\varphi}}+\frac{1}{2}\frac{P_{f}^{2}E^{x\prime}}{E^{\varphi2}E^{x}}+\frac{1}{2}\frac{E^{x\prime}f^{\prime2}}{E^{\varphi2}E^{x}},
\end{equation}
has a remarkable property: its complete form or its vacuum form ($f=0=P_{f}$)
both have a strong Abelian algebra with itself, namely
\begin{align}
\{\mathcal{H}(x),\mathcal{H}(y)\}_{D}= & 0,\\
\left\{ \mathcal{H}(x)\bigg|_{f=0,P_{f}=0},\mathcal{H}(y)\bigg|_{f=0,P_{f}=0}\right\} _{D}= & 0.
\end{align}
This way the algebra of the constraints becomes a Lie algebra since
now we have structure constants instead of structure functions.

\section{Equations of motion\label{sec:Equations-of-motion}}

We are going to write the equations of motion in two equivalent cases,
before and after rescaling lapse and shift and introducing $U_{x}$.
The reason is that to compare the original CGHS equations of motion
with our formulation, it is much easier to use the Hamiltonian equations
of motion before rescaling lapse and shift. This is because in order
to be able to compare equations of motion between our formulation
and the original Lagrangian formulation, we need to find the second
order equations of motion from the Hamiltonian equations of motion.
This is achieved much easier using explicit form of the Hamiltonian
before rescaling.

\subsection{Before introducing $U_{x}$ and rescaling $N$ and $N^{1}$ }

Here we can use the Hamiltonian (\ref{eq:Htot-fin}) and the Dirac
brackets (\ref{eq:Dirac-Br-pairs})-(\ref{eq:Dirac-Br-other}) to
find the the equations of motion, $\dot{F}=\left\{ F(x),\int dyH(y)\right\} _{D}$,
for any function of the phase space $F$. Using these, one gets for
the canonical pairs
\begin{align}
\dot{K}_{x}= & N\Bigg(-\frac{K_{x}K_{\varphi}}{E^{x}}+\frac{1}{2}\frac{f^{\prime2}}{E^{x}E^{\varphi}}+\frac{1}{2}\frac{P_{f}^{2}}{E^{x}E^{\varphi}}+\frac{E^{x\prime\prime}}{E^{x}E^{\varphi}}-\frac{E^{x\prime}E^{\varphi\prime}}{E^{x}E^{\varphi2}}-\frac{E^{x\prime2}}{E^{x2}E^{\varphi}}\Bigg)\nonumber \\
 & +\left(N^{1}K_{x}\right)^{\prime}-\frac{E^{\varphi\prime}N'}{E^{\varphi2}}+{\displaystyle \frac{N''}{E^{\varphi}}},\label{eq:eqmot-Kxdot}\\
\dot{E}_{x}= & NK_{\varphi}+N^{1}E^{x\prime},\label{eq:eqmot-Exdot}\\
\dot{K}_{\varphi}= & N\Bigg(-\frac{1}{2}\frac{f^{\prime2}}{E^{\varphi2}}-\frac{1}{2}\frac{P_{f}^{2}}{E^{\varphi2}}+\frac{1}{2}E^{x}\Lambda+\frac{1}{2}\frac{K_{\varphi}^{2}}{E^{x}}+\frac{1}{2}\frac{E^{x\prime2}}{E^{x}E^{\varphi2}}\Bigg)\nonumber \\
 & +\frac{N'E^{x\prime}}{E^{\varphi2}}+N^{1}K_{\varphi}',\label{eq:eqmot-Kfdot}\\
\dot{E}_{\varphi}= & NK_{x}+N^{1\prime}E^{\varphi}+N^{1}E^{\varphi\prime},\label{eq:eqmot-Efdot}\\
\dot{f}= & \frac{NP_{f}}{E^{\varphi}}+N^{1}f',\label{eq:eqmot-fdot}\\
\dot{P}_{f}= & \left(N\frac{f'}{E^{\varphi}}+N^{1}P_{f}\right)^{\prime}.\label{eq:eqmot-Pfdot}
\end{align}

\subsection{After introducing $U_{x}$ and rescaling $N$ and $N^{1}$\label{sub:eqmot-after}}

In this case, we can use the total Hamiltonian (\ref{eq:HT-deriv-mat})
and the brackets (\ref{eq:brackets-main}) to calculate the equations
of motion, $\dot{F}=\left\{ F(x),\int dyH_{T}(y)\right\} _{D}$. For
the canonical pairs we get: 
\begin{align}
\dot{U}_{x}= & \bar{N}\Bigg(\frac{K_{\varphi}P_{f}f'}{E^{x2}E^{\varphi}}-\frac{f'f''}{E^{x}E^{\varphi2}}-\frac{P_{f}P_{f}^{\prime}}{E^{x}E^{\varphi2}}+\frac{E^{\varphi\prime}(P_{f}^{2}+f^{\prime2})}{E^{x}E^{\varphi3}}\Bigg)\nonumber \\
 & +\bar{N}'\Bigg(-\frac{1}{2}\frac{P_{f}^{2}+f^{\prime2}}{E^{x}E^{\varphi2}}-2\frac{E^{x\prime}E^{\varphi\prime}}{E^{x}E^{\varphi3}}-\frac{1}{2}\frac{K_{\varphi}^{2}}{E^{x2}}+\frac{E^{x\prime\prime}}{E^{x}E^{\varphi2}}-\frac{1}{2}\frac{E^{x\prime2}}{E^{x2}E^{\varphi2}}+2\lambda^{2}\Bigg)\nonumber \\
 & +\bar{N}''\left(\frac{E^{x\prime}}{E^{x}E^{\varphi2}}\right)+\left(\bar{N}^{1}U_{x}\right)^{\prime},\label{eq:eqmot-Uxdot-rscl}\\
\dot{E}_{x}= & \bar{N}^{1}E^{x\prime},\label{eq:eqmot-Exdot-rscl}\\
\dot{K}_{\varphi}= & \bar{N}\Bigg(-\frac{E^{x\prime}\left(P_{f}^{2}+f^{\prime2}\right)}{E^{x}E^{\varphi3}}+\frac{f'P_{f}K_{\varphi}^{2}}{E^{x}E^{\varphi2}}\Bigg)+\bar{N}'\frac{E^{x\prime2}}{E^{x}E^{\varphi3}}+\bar{N}^{1}K_{\varphi}',\label{eq:eqmot-Kfdot-rscl}\\
\dot{E}_{\varphi}= & \bar{N}\frac{P_{f}f'}{E^{x}E^{\varphi}}-\bar{N}'\frac{K_{\varphi}}{E^{x}}+\left(\bar{N}^{1}E^{\varphi}\right)^{\prime},\label{eq:eqmot-Efdot-rscl}\\
\dot{f}= & \bar{N}\left(\frac{P_{f}E^{x\prime}}{E^{x}E^{\varphi2}}-\frac{K_{\varphi}f'}{E^{x}E^{\varphi}}\right)+\bar{N}^{1}f',\label{eq:eqmot-fdot-rscl}\\
\dot{P}_{f}= & \left[\bar{N}\left(\frac{E^{x\prime}f'}{E^{x}E^{\varphi2}}-\frac{K_{\varphi}P_{f}}{E^{x}E^{\varphi}}\right)+\bar{N}^{1}P_{f}\right]^{\prime}.\label{eq:eqmot-Pfdot-rscl}
\end{align}

\section{Gauge fixing the vacuum theory\label{sec:Gauge-fixing-vaccum}}

Using (\ref{eq:HT-deriv-mat}), the vacuum Hamiltonian (with $f=P_{f}=0$)
after an integration by parts can be written as
\begin{align}
H_{0}= & \overline{N}'\mathcal{H}_{d}+\overline{N}^{1}\mathcal{D}\nonumber \\
= & \overline{N}'\left(\frac{1}{2}\frac{E^{x\prime2}}{E^{\varphi2}E^{x}}-2E^{x}\lambda^{2}-\frac{1}{2}\frac{K_{\varphi}^{2}}{E^{x}}\right)+\overline{N}^{1}\left(-U_{x}E^{x\prime}+E^{\varphi}K_{\varphi}'\right),\label{eq:Htot-intpart-nomass}
\end{align}
where the prime sign on $\overline{N}'$ can be taken to mean that
we have a new lapse after integration by parts or to show that this
lapse is the derivative of the previous lapse $\overline{N}$ with
respect to $x$. To completely solve the vacuum classical theory,
one can use two gauge fixings. We choose the first gauge fixing with
an explicit coordinate dependence as
\begin{equation}
\chi_{1}=E^{x}(x)-e^{-2\lambda x}\approx0.\label{eq:GFXEx}
\end{equation}
Since this gauge fixing does not contain any explicit time dependence,
its preservation condition in time gives
\begin{align}
\dot{\chi}_{1}(x)\approx & 0\\
\int dy\{\chi_{1}(x),H_{0}(y)\}_{D}\approx & 0\\
\overline{N}^{1}(x)E^{\prime}(x)\approx & 0,
\end{align}
which implies that the shift (not the original shift but the rescaled
one) vanishes,
\begin{equation}
\overline{N}^{1}=0.\label{eq:N1pres0}
\end{equation}
The motivation for the gauge fixing (\ref{eq:GFXEx}) is the following:
the coordinate transformations
\begin{align}
\overline{x}^{+}= & \frac{e^{\lambda(t+x)}}{\lambda},\label{eq:coor-trans-x-sigma-1}\\
\overline{x}^{-}= & -\frac{e^{-\lambda(t-x)}}{\lambda},\label{eq:coor-trans-x-sigma-2}
\end{align}
will lead us to the formulation of the theory in the conformal gauge
in the original CGHS paper \citep{C.G.Callan1992}, where $\left(\overline{x}^{-},\overline{x}^{+}\right)$
are the null coordinates in that gauge. Now if we use these transformations
in (\ref{eq:GFXEx}), the form of $E^{x}$ will becomes exactly the
same as the form it gets in the vacuum case in the conformal gauge
in \citep{C.G.Callan1992}.

We choose our next gauge fixing as
\begin{equation}
\chi_{2}=E^{\varphi}(x)-1\approx0.\label{eq:GFXEf}
\end{equation}
The reason behind this gauge fixing is the observation that in the
conformal gauge, $E^{\varphi}$ is the only independent metric component
and upon using the coordinate transformations (\ref{eq:coor-trans-x-sigma-1})
and (\ref{eq:coor-trans-x-sigma-2}), the vacuum metric becomes flat
(i.e. $E^{\varphi}$=1). Preserving the $\chi_{2}$ constraint gives
\begin{align}
\dot{\chi}_{2}(x)\approx & 0\\
\int dy\{\chi_{2}(x),H_{1}(y)\}_{D}\approx & 0\\
\overline{N}^{\prime}(x)K_{\varphi}(x)e^{2\lambda x}\approx & 0
\end{align}
which yields
\begin{equation}
\overline{N}^{\prime}(x)=0.\label{eq:Nppres0}
\end{equation}
This is not an issue since this is the derivative of the transformed
original lapse. Up to now, we have $\overline{N}',\,\overline{N}^{1},\, E^{x}$
and $E^{\varphi}$ explicitly. Now we can solve for $K_{\varphi}$
from the weakly vanishing of the Hamiltonian constraint in (\ref{eq:Htot-intpart-nomass}).
This, together with (\ref{eq:GFXEx}) and (\ref{eq:GFXEf}) yields
\begin{equation}
K_{\varphi}=0.\label{eq:KfGFX}
\end{equation}
Finally $U_{x}$ can be found from the weakly vanishing the diffeomorphism
constraint as

\begin{equation}
U_{x}=\frac{E^{\varphi}K_{\varphi}^{\prime}}{E^{x\prime}},\label{eq:Ux-diff-nomass}
\end{equation}
which upon substituting the relevant values of the variables from
above gives
\begin{equation}
U_{x}=0.
\end{equation}
This way, the vacuum case is completely solved classically.

\subsection{The original lapse and shift}

Using the results we just obtained for the vacuum case, we can express
the original lapse and shift in (\ref{eq:HtotfnUx}) in terms of the
ones we obtained using the gauge fixings. Using (\ref{eq:N-redef}),
(\ref{eq:Nppres0}), (\ref{eq:GFXEx}) and (\ref{eq:GFXEf}) we get
\begin{align}
\overline{N}^{\prime}= & 0\\
\left(N\frac{E^{\varphi}E^{x}}{E^{x\prime}}\right)^{\prime}= & 0\\
N^{\prime}= & 0,
\end{align}
which implies
\begin{equation}
N=g(t)+C,
\end{equation}
for which we can simply choose
\begin{equation}
N=1.
\end{equation}
Using (\ref{eq:N1-redef}), (\ref{eq:N1pres0}), (\ref{eq:GFXEx})
and (\ref{eq:KfGFX}) we find
\begin{align}
\overline{N}^{1}= & 0\\
N^{1}(x)+\frac{N(x)K_{\varphi}}{E^{x\prime}}= & 0\\
N^{1}(x)= & 0.
\end{align}

\section{Gauge fixing the case coupled to matter\label{sec:Gauge-fixing-matter}}

The above gauge fixings which we used for the vacuum case will not
work for the case with matter field. Among other reasons, one relatively
obvious reason is that the two gauge fixings (\ref{eq:GFXEx}) and
(\ref{eq:GFXEf}), are related to the form that $E^{x}$ and $E^{\varphi}$
take in the \emph{vacuum} case. Also we can not turn the whole Hamiltonian
constraint into a total derivative and therefore if we get a vanishing
lapse, it would be the original lapse itself that vanishes not its
derivative. By inspecting the total Hamiltonian (\ref{eq:HT-deriv-mat}),
one can see that perhaps a good choice for the first gauge fixing
is
\begin{equation}
\zeta_{1}=E^{x}-h(x)\approx0.\label{eq:GFX-mat-Ex}
\end{equation}
with $h(x)$ an arbitrary function of $x$ coordinate. Since the only
nonvanishing Dirac bracket between $\zeta_{1}$ and $H_{T}$ in (\ref{eq:HT-deriv-mat})
comes from the term in the diffeomorphism constraint containing $U_{x}$,
and since $U_{x}$ appears without derivatives there, the preservation
of $\zeta_{1}$ will give us an algebraic equation for $\bar{N}^{1}$.
In more precise way we have
\begin{align}
\dot{\zeta}_{1}=\left\{ \zeta_{1}(x),\int dyH_{T}(y)\right\}  & \approx0\\
-\int dy\overline{N}^{1}(y)\left\{ E^{x}(x),U_{x}(y)\right\} E^{x\prime}(y) & \approx0\\
\int dy\overline{N}^{1}(y)E^{x\prime}(y)\delta(x-y) & \approx0\\
\overline{N}^{1}(x)E^{x\prime}(x) & \approx0,
\end{align}
which means that the shift vanishes, 
\begin{equation}
\overline{N}^{1}=0.\label{eq:N1-mat-GFX1}
\end{equation}
Now, from (\ref{eq:HT-deriv-mat}), (\ref{eq:GFX-mat-Ex}) and (\ref{eq:N1-mat-GFX1}),
the partially gauge fixed total Hamiltonian will become
\begin{equation}
H_{TF}=\overline{N}\left(\frac{\partial}{\partial x}\left(\frac{1}{2}\frac{h^{\prime2}}{hE^{\varphi2}}-2h\lambda^{2}-\frac{1}{2}\frac{K_{\varphi}^{2}}{h}\right)-\frac{f^{\prime}P_{f}K_{\varphi}}{hE^{\varphi}}+\frac{1}{2}\frac{P_{f}^{2}h'}{hE^{\varphi2}}+\frac{1}{2}\frac{h'f^{\prime2}}{hE^{\varphi2}}\right).\label{eq:HT-deriv-mat-GF1}
\end{equation}
For the second gauge fixing, we follow the procedure suggested in
\citep{Gambini2013,Alvarez2012}. The basic idea is to make a canonical
transformation and define a true Hamiltonian that gives the correct
equations of motion. To start, we identify the terms inside the total
derivative of the Hamiltonian constraint in (\ref{eq:HT-deriv-mat-GF1})
as our new canonical variable $X$,
\begin{equation}
X=\frac{1}{2}\frac{h^{\prime2}}{hE^{\varphi2}}-2h\lambda^{2}-\frac{1}{2}\frac{K_{\varphi}^{2}}{h}.\label{eq:Def-X}
\end{equation}
This suggests that we can find a generating function of the third
kind as
\begin{equation}
F_{3}(p,Q)=F_{3}(K_{\varphi},X),
\end{equation}
by means of the equation
\begin{equation}
E^{\varphi}(X,K_{\varphi})=\frac{\partial F_{3}(K_{\varphi},X)}{\partial K_{\varphi}}.
\end{equation}
One can find $E^{\varphi}$ from (\ref{eq:Def-X}), substitute it
into the above and integrate with respect to $K_{\varphi}$ to get
\begin{equation}
F_{3}(K_{\varphi},X)=h'\ln\left(K_{\varphi}+\Omega\right),
\end{equation}
with
\begin{equation}
\Omega=\sqrt{K_{\varphi}^{2}+2hX+4\lambda^{2}h^{2}}.
\end{equation}
Now $P_{X}$, the momentum conjugate to $X$, can be found as
\begin{equation}
P_{X}=-\frac{\partial F_{3}(K_{\varphi},X)}{\partial X}=-\frac{hh'}{\Omega\left(K_{\varphi}+\Omega\right)}.\label{eq:Def-Px}
\end{equation}
The above equation will be the Hamiltonian constraint after writing
$K_{\varphi}$ in terms of $X,\, f$ and $P_{f}$. To do this, we
find $E^{\varphi}$ from (\ref{eq:Def-X}), substitute it in (\ref{eq:HT-deriv-mat-GF1}),
find $K_{\varphi}$ from vanishing of it and substitute the resulting
$K_{\varphi}$ into (\ref{eq:Def-Px}). This way, our total Hamiltonian
will be
\begin{equation}
H_{\textrm{tot}}=\bar{N}P_{X}+H_{\textrm{true}}=\bar{N}\left(P_{X}+\frac{hh'}{\Omega(X,f,P_{f})\left(K_{\varphi}(X,f,P_{f})+\Omega(X,f,P_{f})\right)}\right).\label{eq:HT-mat-GFX2}
\end{equation}
The next step is to introduce the second gauge condition as
\begin{equation}
\zeta_{2}=X-b(x,t)\approx0,
\end{equation}
with $b(x,t)$ a function of the coordinates. Since in (\ref{eq:HT-mat-GFX2})
$P_{X}$ only appears in the first term in the parenthesis, the preservation
of the above constraint
\begin{equation}
\dot{\zeta}_{2}=\left\{ \zeta_{2}(x),\int dyH_{\textrm{tot}}(y)\right\} +\frac{\partial b(x,t)}{\partial t}\approx0
\end{equation}
gives
\begin{equation}
\bar{N}=\dot{b}.
\end{equation}
Because $f$ and $P_{f}$ commute with $P_{X}$, the evolution equations
will be
\begin{align}
\dot{f}= & \left\{ f(x),\int dyH_{\textrm{tot}}(y)\right\} =\left\{ f(x),\int dyH_{\textrm{true}}(y)\right\} ,\label{eq:fdot-true}\\
\dot{P}_{f}= & \left\{ P_{f}(x),\int dyH_{\textrm{tot}}(y)\right\} =\left\{ P_{f}(x),\int dyH_{\textrm{true}}(y)\right\} ,\label{eq:Pfdot-true}
\end{align}
with the true Hamiltonian being
\begin{equation}
H_{\textrm{true}}=\dot{b}\left(\frac{hh'}{\Omega(f,P_{f})\left(K_{\varphi}(f,P_{f})+\Omega(f,P_{f})\right)}\right).\label{eq:H-true}
\end{equation}
The true Hamiltonian (\ref{eq:H-true}) is a local Hamiltonian density
in the sense that $\bar{N}$ has not been given in terms of an integral
of canonical variables. This happens thanks to the present method
of gauge fixing which gives an algebraic equation for lapse instead
of a differential equation. If $\bar{N}$ was given in terms of an
integral of canonical variables, then the Hamiltonian would have been
given in terms of an integral of an integral and would have been nonlocal
in that sense. 

This local true Hamiltonian gives the correct equations of motion
for our system as it can be checked by comparing the equations (\ref{eq:fdot-true})
and (\ref{eq:Pfdot-true}) with the equations of motion derived before
gauge fixing in section (\ref{sub:eqmot-after}), and then substituting
in them the gauge fixing conditions.

\section{Comparing with the original Lagrangian theory\label{sec:Comparing-with-original}}

In order to make a connection with the original Lagrangian formulation
of the CGHS and also to check the consistency of our formulation,
we transform our equations of motion into the ones in the null coordinates
in the conformal gauge and compare them to the original Lagrangian
ones.

\subsection{Metric and other variables in null coordinates}

The first step to make a connection between the two formalisms is
finding the relations between the form of the metric and other canonical
variables in both formulations. The coordinates used in the original
CGHS formulation are the null coordinates 
\begin{align}
x^{+}= & x^{0}+x^{1},\label{eq:xplus}\\
x^{-}= & x^{0}-x^{1},\label{eq:xminus}
\end{align}
where $x^{0}=t$ and $x^{1}=x$ are the coordinates used here up to
now. We can find the relation between the components of the null and
non-null metrics using the general transformation 
\begin{equation}
g_{ab}=\frac{\partial x^{a'}}{\partial x^{a}}\frac{\partial x^{b'}}{\partial x^{b}}\bar{g}_{a'b'}.
\end{equation}
In the original CGHS model, the metric components in the conformal
gauge are
\begin{align}
\bar{g}_{+-} & =-\frac{1}{2}e^{2\rho},\label{eq:g+-}\\
\bar{g}_{--} & =\bar{g}_{++}=0.\label{eq:g++}
\end{align}
Thus the relations between the components in two coordinate systems
are 
\begin{align}
g_{00}= & 2\bar{g}_{+-}=-e^{2\rho}.\label{eq:g00null}\\
g_{11}= & -2\bar{g}_{+-}=e^{2\rho}.\label{eq:g11null}\\
g_{01}= & g_{10}=g_{++}-g_{--}=0.\label{eq:g01null}
\end{align}
The relations between the partial derivatives in the two coordinates
thus become 
\begin{align}
\partial_{t}= & \partial_{+}+\partial_{-},\label{eq:d0CGHS}\\
\partial_{x}= & \partial_{+}-\partial_{-},\label{eq:d1CGHS}\\
\partial_{t}\partial_{t}= & \partial_{+}\partial_{+}+\partial_{-}\partial_{-}+2\partial_{+}\partial_{-},\label{eq:d0d0CGHS}\\
\partial_{x}\partial_{x}= & \partial_{+}\partial_{+}+\partial_{-}\partial_{-}-2\partial_{+}\partial_{-},\label{eq:d1d1CGHS}\\
\partial_{t}\partial_{x}= & \partial_{+}\partial_{+}-\partial_{-}\partial_{-}.\label{eq:d0d1CGHS}
\end{align}
We can also write $E^{x}$ in terms of the dilaton field. Using the
relation between $\Phi$ and $\phi$ (the dilaton field in original
CGHS paper) 
\begin{equation}
\Phi=2\sqrt{2}e^{-\phi}.
\end{equation}
and substituting it into the $\mu$ constraint (\ref{eq:muconst})
and equating this second class constraint strongly to zero, we get
\begin{equation}
E^{x}=\frac{1}{4}\Phi^{2}=2e^{-2\phi}.\label{eq:Ex-Phi}
\end{equation}
The variable $E^{\varphi}$ can also be written as 
\begin{align}
E^{\varphi}= & \frac{|P|}{2}=\sqrt{q}=\sqrt{q_{11}}=\sqrt{g_{11}}=\sqrt{-2g_{+-}}=\sqrt{e^{2\rho}}=e^{\rho},\label{eq:Ef-rho}
\end{align}
where we have used the canonical transformation (\ref{eq:Pnorm-Ash})
and the fact that $q_{ab}$ has only one independent component so
that $q=q_{11}$. The details of these calculations can be found in
\citep{us-new-var}. Using the ADM formalism, the form of the metric
components in 2D can be expressed as 
\begin{align}
g_{00}= & -N^{2}+q_{11}(N^{1})^{2},\\
g_{11}= & q_{11},\\
g_{01}= & -q_{11}N^{1}.
\end{align}
From this, one can find $N$ and $N^{1}$ in terms of metric components
as 
\begin{align}
N^{1}= & -\frac{g_{01}}{q_{11}}=-\frac{g_{01}}{g_{11}},\label{eq:metr-ls-1}\\
N= & \sqrt{q_{11}(N^{1})^{2}-g_{00}}=\sqrt{\frac{g_{01}^{2}}{q_{11}}-g_{00}}=\sqrt{\frac{g_{01}^{2}}{g_{11}}-g_{00}}.\label{eq:metr-ls-2}
\end{align}
Substituting (\ref{eq:g00null})-(\ref{eq:g01null}) in the above
two equations and using (\ref{eq:Ef-rho}) yields 
\begin{align}
N^{1} & =0,\label{eq:N1-CGHS}\\
N & =\sqrt{-g_{00}}=e^{\rho}=E^{\varphi}.\label{eq:N-CGHS}
\end{align}
Now we are ready to compare the equations of motion.

\subsection{The equations of motion in null coordinates}

In order to compare our equations of motion with the original ones
in the CGHS paper, we need to transform ours into the second order
form and then bring them into the null form. Starting from the equations
for the matter field and its conjugate, if we find $P_{f}$ from (\ref{eq:eqmot-fdot}),
substitute it into (\ref{eq:eqmot-Pfdot}) and then use (\ref{eq:d0CGHS})-(\ref{eq:d0d1CGHS}),
(\ref{eq:Ef-rho}), (\ref{eq:N1-CGHS}) and (\ref{eq:N-CGHS}) we
get 
\begin{equation}
\partial_{+}\partial_{-}f=0.\label{eq:eqmotnull-matt}
\end{equation}
Finding $K_{\varphi}$ from (\ref{eq:eqmot-Exdot}) and substituting
it in (\ref{eq:eqmot-Kfdot}) and using (\ref{eq:d0CGHS})-(\ref{eq:d0d1CGHS}),
(\ref{eq:Ex-Phi}), (\ref{eq:Ef-rho}), (\ref{eq:N1-CGHS}) and (\ref{eq:N-CGHS})
yields 
\begin{align}
V_{1}=e^{-\rho}e^{-2\phi}\big( & e^{2\phi}[(\partial_{+}f)^{2}+(\partial_{-}f)^{2}]+4e^{2\rho}\lambda^{2}-4\partial_{+}^{2}\phi-4\partial_{-}^{2}\phi\nonumber \\
 & -8\partial_{+}\partial_{-}\phi+16\partial_{+}\phi\partial_{-}\phi+8\partial_{+}\rho\partial_{+}\phi+8\partial_{-}\rho\partial_{-}\phi\big)=0.\label{eq:eqmotnull-ExKf}
\end{align}
For the next second order equation, we find $K_{x}$ from (\ref{eq:eqmot-Efdot})
and substitute it in (\ref{eq:eqmot-Kxdot}). Then upon using (\ref{eq:d0CGHS})-(\ref{eq:d0d1CGHS}),
(\ref{eq:Ex-Phi}), (\ref{eq:Ef-rho}), (\ref{eq:N1-CGHS}), (\ref{eq:N-CGHS})
and the values of $P_{f}$ and $K_{\varphi}$ from (\ref{eq:eqmot-fdot})
and (\ref{eq:eqmot-Exdot}) respectively, we get 
\begin{align}
V_{2}= & -\frac{1}{2}e^{2\phi}[(\partial_{+}f)^{2}+(\partial_{-}f)^{2}]+2\partial_{+}^{2}\phi+2\partial_{-}^{2}\phi-4\partial_{+}\partial_{-}\phi\nonumber \\
 & +4\partial_{+}\partial_{-}\rho-4\partial_{+}\rho\partial_{+}\phi-4\partial_{-}\rho\partial_{-}\phi=0.\label{eq:eqmotnull-EfKx}
\end{align}
We can follow the same procedure and find the Hamiltonian and diffeomorphism
constraints in (\ref{eq:Htot-fin}) in the null coordinate as 
\begin{align}
\mathcal{H}=e^{-\rho}e^{-2\phi}\big( & e^{2\phi}[(\partial_{+}f)^{2}+(\partial_{-}f)^{2}]-4e^{2\rho}\lambda^{2}-4\partial_{+}^{2}\phi-4\partial_{-}^{2}\phi\nonumber \\
 & +8\partial_{+}\partial_{-}\phi-16\partial_{+}\phi\partial_{-}\phi+8\partial_{+}\rho\partial_{+}\phi+8\partial_{-}\rho\partial_{-}\phi\big)=0,\label{eq:eqmotnull-H}\\
\mathcal{D}=e^{-2\phi}\big(8\partial_{+} & \rho\partial_{+}\phi-8\partial_{-}\rho\partial_{-}\phi-4\partial_{+}^{2}\phi+4\partial_{-}^{2}\phi\big)+(\partial_{+}f)^{2}-(\partial_{-}f)^{2}=0.\label{eq:eqmotnull-D}
\end{align}

\subsection{Identifying our equations of motion with those of the CGHS paper}

If we compare the above equations with the equations of motion of
the original CGHS model, we can note the following: 

The matter field equation is identically the same in both methods
and is given by (\ref{eq:eqmotnull-matt}). The original energy momentum
equations, $T_{++}=0$ and $T_{--}=0$ in \citep{C.G.Callan1992}
are combined in the diffeomorphism constraint (\ref{eq:eqmotnull-D})
as 
\begin{align}
T_{++}-T_{--}= & \frac{1}{2}\mathcal{D}\nonumber \\
= & \left[e^{-2\phi}(4\partial_{+}\rho\partial_{+}\phi-2\partial_{+}^{2}\phi)+\frac{1}{2}(\partial_{+}f)^{2}\right]\nonumber \\
 & -\left[e^{-2\phi}(4\partial_{-}\rho\partial_{-}\phi-2\partial_{-}^{2}\phi)+\frac{1}{2}(\partial_{-}f)^{2}\right]\nonumber \\
= & 0.
\end{align}
The original $T_{+-}=0$ equation can be obtained by combining the
Hamiltonian constraint equation (\ref{eq:eqmotnull-H}) and equation
(\ref{eq:eqmotnull-ExKf}) as following: 
\begin{equation}
T_{+-}=-\frac{e^{\rho}}{8}\left(V_{1}-\mathcal{H}\right)=e^{-2\phi}(2\partial_{+}\partial_{-}\phi-4\partial_{+}\phi\partial_{-}\phi-\lambda^{2}e^{2\rho})=0,
\end{equation}
and finally the dilaton field equation of motion in CGHS paper is
obtained by combining $V_{1}$ and $V_{2}$ i.e. equations (\ref{eq:eqmotnull-ExKf})
and (\ref{eq:eqmotnull-EfKx}) as
\begin{equation}
\frac{e^{\rho}e^{2\phi}}{4}V_{1}+\frac{1}{2}V_{2}=-4\partial_{+}\partial_{-}\phi+4\partial_{+}\phi\partial_{-}\phi+2\partial_{+}\partial_{-}\rho+\lambda^{2}e^{2\rho}=0.
\end{equation}

\section{Boundary terms\label{sec:Boundary-terms}}

It is important to take care of the boundary conditions in our theory
\citep{Regge1974}. One important reason is that energy in general
relativity is related to the surface integral or boundary term at
infinity %
\footnote{More precisely it is identified as the conserved quantity associated
to the invariance of the action under time translations at infinity,
i.e. under transformation generated by a timelike killing vector field
at infinity.%
}. 

We require that both Hamiltonian and diffeomorphism constraints
\begin{align}
H_{c}= & \int dxN\mathcal{H},\\
D_{c}= & \int dxN^{1}\mathcal{D},
\end{align}
be functionally differentiable. This means that if the variations
of $H_{c}$ and $D_{c}$ leads to
\begin{align}
\delta H_{c}= & \int_{-\infty}^{\infty}dxN\delta\mathcal{H}=\int_{-\infty}^{\infty}dx\left(\frac{\delta H_{c}}{\delta q^{i}}\delta q^{i}+\frac{\delta H_{c}}{\delta p^{i}}\delta p^{i}\right)+\int_{-\infty}^{\infty}dx\partial_{x}\left(\delta S_{H}\right),\\
\delta D_{c}= & \int_{-\infty}^{\infty}dxN^{1}\delta\mathcal{D}=\int_{-\infty}^{\infty}dx\left(\frac{\delta D_{c}}{\delta q^{i}}\delta q^{i}+\frac{\delta D_{c}}{\delta p^{i}}\delta p^{i}\right)+\int_{-\infty}^{\infty}dx\partial_{x}\left(\delta S_{D}\right),
\end{align}
then for $H_{c}$ and $D_{c}$ to be functionally differentiable,
we need to add $-\delta S_{H}$ and $-\delta S_{D}$ to them respectively.
This means that the overall variation of surface term that should
be added to the variation of the action is 
\begin{equation}
\delta S_{\textnormal{surface}}=-\delta\int dt\left(S_{H}+S_{D}\right),
\end{equation}
and clearly the surface term to be added to the action will be
\begin{equation}
S_{\textnormal{surface}}=-\int dt\left(S_{H}+S_{D}\right).
\end{equation}
In the language of formulation of the CGHS that has been presented
so far, i.e. using the total Hamiltonian (\ref{eq:HT-deriv-mat}),
the terms that obstruct functional differentiability for the diffeomorphism
constraint turn out to be
\begin{equation}
\partial_{x}\left(\delta S_{D}\right)=\partial_{x}\left[\bar{N}^{1}\left(E^{\varphi}\delta K_{\varphi}+P_{f}\delta f-U_{x}\delta E^{x}\right)\right]
\end{equation}
and the corresponding terms for Hamiltonian constraint are
\begin{align}
\partial_{x}\left(\delta S_{H}\right)=\partial_{x}\bigg[\bar{N}\bigg( & \frac{1}{2}\frac{K_{\varphi}^{2}}{E^{x2}}\delta E^{x}+\frac{E^{x\prime\prime}\delta E^{x}}{E^{\varphi2}E^{x}}+\frac{1}{2}\frac{f^{\prime2}\delta E^{x}}{E^{\varphi2}E^{x}}+\frac{E^{x\prime}f'\delta f}{E^{\varphi2}E^{x}}\nonumber \\
 & -2\lambda^{2}\delta E^{x}-\frac{E^{x\prime2}\delta E^{\varphi}}{E^{\varphi3}E^{x}}-2\frac{E^{\varphi\prime}E^{x\prime}\delta E^{x}}{E^{\varphi3}E^{x}}-\frac{K_{\varphi}\delta K_{\varphi}}{E^{x}}\nonumber \\
 & +\frac{3}{2}\frac{E^{x\prime2}\delta E^{x}}{E^{\varphi2}E^{x2}}-\frac{K_{\varphi}P_{f}\delta f}{E^{\varphi}E^{x}}+\frac{1}{2}\frac{P_{f}^{2}\delta E^{x}}{E^{\varphi2}E^{x}}\bigg)\bigg]\nonumber \\
+\partial_{x}\bigg[\partial_{x} & \left(\frac{\bar{N}E^{x\prime}}{E^{\varphi2}E^{x}}\delta E^{x}\right)\bigg].
\end{align}
So the total variation of the surface term is
\begin{align}
\delta S_{\textnormal{surface}}=-\int dt & \left(\int_{-\infty}^{\infty}dx\partial_{x}\left(\delta S_{H}\right)+\int_{-\infty}^{\infty}dx\partial_{x}\left(\delta S_{H}\right)\right)\nonumber \\
=-\int dt & \left[\delta S_{H}+\delta S_{D}\right]_{-\infty}^{\infty}\nonumber \\
=-\int dt & \bigg\{\bar{N}^{1}\left(E^{\varphi}\delta K_{\varphi}+P_{f}\delta f-U_{x}\delta E^{x}\right)\nonumber \\
 & +\bar{N}\bigg[\bigg(\frac{1}{2}\frac{K_{\varphi}^{2}}{E^{x2}}+\frac{E^{x\prime\prime}}{E^{\varphi2}E^{x}}+\frac{1}{2}\frac{f^{\prime2}}{E^{\varphi2}E^{x}}-2\lambda^{2}\nonumber \\
 & -2\frac{E^{\varphi\prime}E^{x\prime}}{E^{\varphi3}E^{x}}+\frac{3}{2}\frac{E^{x\prime2}}{E^{\varphi2}E^{x2}}+\frac{1}{2}\frac{P_{f}^{2}}{E^{\varphi2}E^{x}}\bigg)\delta E^{x}\nonumber \\
 & +\left(\frac{E^{x\prime}f'}{E^{\varphi2}E^{x}}-\frac{K_{\varphi}P_{f}}{E^{\varphi}E^{x}}\right)\delta f-\frac{E^{x\prime2}\delta E^{\varphi}}{E^{\varphi3}E^{x}}-\frac{K_{\varphi}\delta K_{\varphi}}{E^{x}}\bigg]\nonumber \\
 & +\partial_{x}\left(\frac{\bar{N}E^{x\prime}}{E^{\varphi2}E^{x}}\delta E^{x}\right)\bigg\}_{-\infty}^{\infty}.\label{eq:delta-surf-gen}
\end{align}
If we use the prescription at infinity for the matter field 
\begin{equation}
\delta f\big|_{x=\pm\infty}=0,\label{eq:usual-presc}
\end{equation}
then the variation of the surface term will be
\begin{align}
\delta S_{\textnormal{surface}}=-\int dt & \bigg\{\bar{N}^{1}\left(E^{\varphi}\delta K_{\varphi}-U_{x}\delta E^{x}\right)\nonumber \\
 & +\bar{N}\bigg[\bigg(\frac{1}{2}\frac{K_{\varphi}^{2}}{E^{x2}}+\frac{E^{x\prime\prime}}{E^{\varphi2}E^{x}}+\frac{1}{2}\frac{f^{\prime2}}{E^{\varphi2}E^{x}}-2\lambda^{2}\nonumber \\
 & -2\frac{E^{\varphi\prime}E^{x\prime}}{E^{\varphi3}E^{x}}+\frac{3}{2}\frac{E^{x\prime2}}{E^{\varphi2}E^{x2}}+\frac{1}{2}\frac{P_{f}^{2}}{E^{\varphi2}E^{x}}\bigg)\delta E^{x}\nonumber \\
 & -\frac{E^{x\prime2}\delta E^{\varphi}}{E^{\varphi3}E^{x}}-\frac{K_{\varphi}\delta K_{\varphi}}{E^{x}}\bigg]+\partial_{x}\left(\frac{\bar{N}E^{x\prime}}{E^{\varphi2}E^{x}}\delta E^{x}\right)\bigg\}_{-\infty}^{\infty}.\label{eq:delta-surf-gen-presc}
\end{align}
For the present gauge fixing, one can arrive at a surface term by
substituting the gauge fixings and the values of lapse and shift into
the general form of the variation of the total surface term (\ref{eq:delta-surf-gen-presc}),
and also putting $\delta h(x)=0=\delta b(x,t)$ (since they are just
functions of the coordinates):
\begin{equation}
S_{\textnormal{surface}}=\int dt\left[\frac{1}{2}\frac{\dot{b}}{h}\left(K_{\varphi}^{2}-\frac{h^{\prime2}}{E^{\varphi2}}\right)\right]_{-\infty}^{\infty}.
\end{equation}
Comparing the above term to the standard form of the boundary term
of the CGHS \citep{Kiefer2012},
\[
S_{\textrm{surface}}=-\int dt\left[NM\right]_{-\infty}^{\infty},
\]
where $M$ is the ADM mass, and since $\bar{N}=\dot{b}$ by our gauge
fixing, one gets
\[
M=\frac{1}{2h}\left(\frac{h^{\prime2}}{E^{\varphi2}}-K_{\varphi}^{2}\right).
\]
By using (\ref{eq:Def-X}), the above formula can also be written
as 
\begin{align*}
M= & X+2h\lambda^{2}\\
= & b+2h\lambda^{2}
\end{align*}
where we used the second gauge fixing in the second line.

\section{Conclusion}

We have analyzed the CGHS model without any conformal transformation
in terms of new variables similar to the standard Ashtekar variables
of the 3+1 spherically symmetric model. This means that not only the
variables have direct geometric meaning and there is no need to turn
everything back to their directly-geometric form at the end, but also
it might be much easier to read off the physics out of the this formulation.%
{} Then, by means of rescaling lapse and shift, the Hamiltonian constraint
of the system was cast into a form such that it commutes with itself
both in vacuum and coupled-to-matter cases. This makes the system
suitable for analysis using loop quantum gravity techniques because,
among other things, the system is expressed in Ashtekar-like variables
and the algebra of constraints is now a Lie algebra. In the next step,
we solved both vacuum and coupled-to-matter cases completely classically
by introducing two gauge fixings for each case. In the case where
there is matter coupling, we arrive at a local true Hamiltonian hence
leading to local equations of motion. %

As one possible future direction, in not-gauge-fixed or partially
gauge fixed case, it is desirable to polymerize the relevant variables
that captures the semiclassical behavior of loop quantum gravity.
In the totally gauge fixed case, one can follow different quantization
schemes suitable for a local true Hamiltonian theory. These can be
starting points for addressing the issue of singularity resolution
in further works. Also the possibility of completing the Dirac procedure
is in principle there, since the constraint algebra is now a Lie algebra.
In addition, it is also worth noting that the algebra of Hamiltonian
constraint with itself is very simple and this might be a good aid
in a possible process of quantization.
\begin{acknowledgments}
I would like to thank Rodolfo Gambini, Jorge Pullin, Alejandro Corichi,
Tatjana Vukasinac and Asieh Karami for discussions, guidance and corrections.
This work was supported by the grant from the Programa de Becas Posdoctorales,
Centro de Ciencias Matematicas, Campos Morelia, UNAM.
\end{acknowledgments}
\bibliographystyle{apsrev4-1}
\bibliography{CMass}

\end{document}